\documentclass[aps, prb, twocolumn, superscriptaddress]{revtex4-1}

\usepackage{graphicx}
\usepackage{gensymb}
\usepackage{esvect}
\usepackage{multirow}

\begin{document}

\title{Atomically flat two-dimensional silicon crystals with versatile electronic properties}

\author{Kisung Chae}
\affiliation{Korea Institute for Advanced Study, Seoul 02455, South Korea}
\author{Duck Young Kim}
\affiliation{Center for High Pressure Science and Technology Advanced Research, Shanghai 201203, P. R. China}
\author{Young-Woo Son}
\affiliation{Korea Institute for Advanced Study, Seoul 02455, South Korea}
\email{hand@kias.re.kr}

\date{\today}

\begin{abstract}
Silicon (Si) is one of the most extensively studied materials owing to its significance to semiconductor science and technology. While efforts to find a new three-dimensional (3D) Si crystal with unusual properties have made some progress, its two-dimensional (2D) phases have not yet been explored as much. Here, based on a newly developed systematic \emph{ab initio} materials searching strategy, we report a series of novel 2D Si crystals with unprecedented structural and electronic properties. The new structures exhibit perfectly planar outermost surface layers of a distorted hexagonal network with their thicknesses varying with the atomic arrangement inside. Dramatic changes in electronic properties ranging from semimetal to semiconducting with indirect energy gaps and even to one with direct energy gaps are realized by varying thickness as well as by surface oxidation. Our predicted 2D Si crystals with flat surfaces and tunable electronic properties will shed light on the development of silicon-based 2D electronics technology.
\end{abstract}

\maketitle

\section{Introduction}

Recently, various 2D materials with weak van der Waals (vdW) interlayer interaction have been extensively studied due to their unusual properties~\cite{geim_van_2013,novoselov_two-dimensional_2005,lee_atomically_2014}. Examples of these include graphene, hexagonal boron nitride, black phosphorous, and transition metal dichalcogenides. Not only that they have shown superior physical and chemical properties, some of the models in theoretical physics such as massless Dirac fermions have also been realized in experiments, which otherwise have not been observed in conventional materials~\cite{geim_van_2013}. Nevertheless, many practical issues about large-scale synthesis, processing for defects, and contaminant control still need to be resolved~\cite{geim_van_2013,lin_defect_2016}, which are critical for them to be realized as next generation electronic devices and energy applications.

Silicon, on the other hand, has served as a mainstay of semiconductor technologies, and a vast amount of advanced processing techniques have been accumulated for decades. It is mainly due to its abundance on the Earth surface as well as the existence of a single oxide form (SiO$_2$) which is advantageous to the mass production of a single-element device that is free from phase separations. These make undoubtedly Si be unique in current semiconductor technologies. Therefore, despite very active researches on the aforementioned 2D materials as a next generation platform for various applications, the best candidate may still be Si itself. This leads us to believe that discovering a novel 2D phase of Si materials with desirable physical properties would be important. Compared with the number of efforts for new bulk phase of Si~\cite{wentorf_two_1963,kasper_clathrate_1965,besson_electrical_1987,von_schnering_lithium_1988,gryko_low-density_2000,malone_ab_2008,kim_synthesis_2015,rapp_experimental_2015,botti_low-energy_2012,wang_direct_2014,lee_computational_2014,lee_ab_2016,guo_new_2015,luo_si10:_2016,liu_new_2017}, however, searching for a new 2D Si crystalline phase has not been remarkably succeeded yet, and only a few theoretical predictions~\cite{morishita_formation_2008,bai_graphene-like_2010,morishita_first-principles_2010,morishita_surface_2011,spencer_reconstruction_2012,guo_structural_2015,sakai_structural_2015,aierken_first-principles_2016} and experimental reports~\cite{nakano_preparation_2005,nakano_soft_2006,kim_synthesis_2011,lu_synthesis_2011,kim_scalable_2014,ohsuna_monolayer--bilayer_2016} exist in the literature.

Silicene~\cite{takeda_theoretical_1994,guzman-verri_electronic_2007,cahangirov_two-_2009,vogt_silicene:_2012,tao_silicene_2015,le_lay_2d_2015}, a monolayer form of the 2D Si crystals which is analogous to graphene, cannot form a stable layered structure by itself since surface of the silicene is chemically reactive, so that the adjacent silicene layers form strong covalent bonds~\cite{geim_van_2013,vogt_silicene:_2012}. This is due to the strong preference of Si for the sp$^3$ hybridization over the sp$^2$ in contrast to carbon with the same number of valence electrons. Thus, silicene may not be a good candidate for a scalable 2D phase of Si~\cite{geim_van_2013}. In addition, due to the strong covalent bonding character of Si, as-cleaved surfaces inevitably have unpaired electrons localized at dangling bonds on the surface, which makes an energetically unfavorable situation. This is evidenced by prevalent severe surface reconstruction to reduce the number of unpaired electrons as can be seen in most of the previously reported 2D Si crystals~\cite{bai_graphene-like_2010,morishita_first-principles_2010,morishita_surface_2011,spencer_reconstruction_2012,guo_structural_2015,sakai_structural_2015,aierken_first-principles_2016,ohsuna_monolayer--bilayer_2016}. In all the cases, however, some of the surface atoms still remain under-coordinated even after the reconstruction, implying that those atoms prone to form strong covalent bonding with one another as pointed out in the case of silicene~\cite{geim_van_2013,vogt_silicene:_2012}.

In this work, we theoretically predict a series of novel 2D allotrope of Si crystals constructed by a concrete \emph{ab initio} materials search strategy. The predicted 2D crystals show characteristic structural features as follows. The crystal is composed of two parts: (1) the atomically flat surface layers and (2) the inner layer connecting them through sp$^3$-like covalent bonds as seen in FIG.~\ref{fig1}. The surface layer features perfectly planar stable hexagonal framework unlike other 2D Si crystals that have hitherto been studied~\cite{bai_graphene-like_2010,morishita_first-principles_2010,morishita_surface_2011,spencer_reconstruction_2012,guo_structural_2015,sakai_structural_2015,aierken_first-principles_2016,nakano_preparation_2005,nakano_soft_2006,kim_synthesis_2011,lu_synthesis_2011,kim_scalable_2014,ohsuna_monolayer--bilayer_2016}, in which buckled surfaces are predominant. Moreover, the crystal is completely free from coordination number (CN) defects. The structures composed of two parts are revealed stable against serious perturbations as will be discussed later.

\section{Computational Details}

We concisely describe a novel 2D crystal structure prediction method: namely, Search by \emph{Ab initio} Novel Design \emph{via} Wyckoff position Iterations in the Conformational Hypersurface (abbreviated as \texttt{SANDWICH}), which explores the conformational hyperspace to find various local minima systematically. The method is especially suited for predicting 2D phases of covalent materials by designing the 2D crystals free of CN defects. This can be achieved by building surface and inner parts with different symmetries from each other, and by joining the two parts in such a way that under-coordinated atoms at the interface are compensated by one another. By doing so, the crystal becomes stabilized by eliminating dangling bonds.

Specifically, we chose surface layers to have a space group of P6/mmm (No. 191), while the bulk maintains a sp$^3$ bonding character. Among special positions in the given space group, we find that Wyckoff sites of $e$ (0, 0, $\pm$z) with a point group of 6mm and $i$ [(1/2, 0, $\pm$z), (0, 1/2, $\pm$z) and (1/2, 1/2, $\pm$z)] with the group of 2mm are suitable for building the CN defect-free crystals. Also, we considered two relative positions of the surface layers in this study, represented by displacement vectors of $\vec{d}$=$\mathbf{0}$ and $\vec{d}$ = 0.5 $\vec{a}_{1}$ + 0.5 $\vec{a}_{2}$, where $\vec{a}_1$ and $\vec{a}_2$ are lattice vectors. With all the settings described so far, we generated structures by varying the number of atoms in the inner layer ($n$) consecutively from 0 up to 9. Our method exhausts all the possible combinations for atomic positions of 2D crystals with given constraints: surface symmetry, choice of Wyckoff positions and thickness as schematically shown in FIG.~S1. We note that the method is undoubtedly advantageous since it explores almost all the structures from a highly probable subset of the entire search space for given conditions. More detailed descriptions of our \texttt{SANDWICH} method can be found in the supplementary material.

We performed a series of first-principles calculations to obtain optimized structures by using Vienna \emph{ab initio} simulation package (\texttt{VASP}) code~\cite{kresse_efficient_1996,kresse_efficiency_1996}. Conjugate gradient method was used to find the equilibrium structures with a force criterion of 1 meV/\AA. For Kohn-Sham orbital, core part was treated by using projector augmented wave method~\cite{kresse_ultrasoft_1999}, while the valence part was approximated by linear expansion of a plane wave basis set with the kinetic energy cutoff of 450 eV. Self-consistent field of DFT was iterated until the differences of the total energy and eigenvalues are less than 10$^{-7}$ eV. Numerical integration in the first Brillouin zone (BZ) was done on the $\Gamma$-centered 12$\times$12$\times$1 grid meshes generated by a Monkhorst-Pack scheme. The exchange-correlation functional of Perdew-Burke-Ernzerhof was used to build the Hamiltonian of an electron-ion system~\cite{perdew_generalized_1996}. For a better description of electronic structures with a band gap, a hybrid functional of Heyd-Scuseria-Ernzerhof~\cite{krukau_influence_2006} as implemented in the \texttt{VASP} code was used. Dynamical stability was also checked on each of the relaxed structures. Phonon dispersion spectra were generated by using a direct method~\cite{parlinski_first-principles_1997} as implemented in \texttt{phonopy} package~\cite{togo_first_2015}. To obtain force constants, we used 4$\times$4$\times$1 and 5$\times$5$\times$1 supercells to generate displaced configurations. In this case, the k-point sampling in the BZ was done on the 24$\times$24$\times$1 and 25$\times$25$\times$1 Monkhorst-Pack grids.

\section{Results and Discussion}

\begin{figure}[b]
  \centering
  \includegraphics[width=\columnwidth]{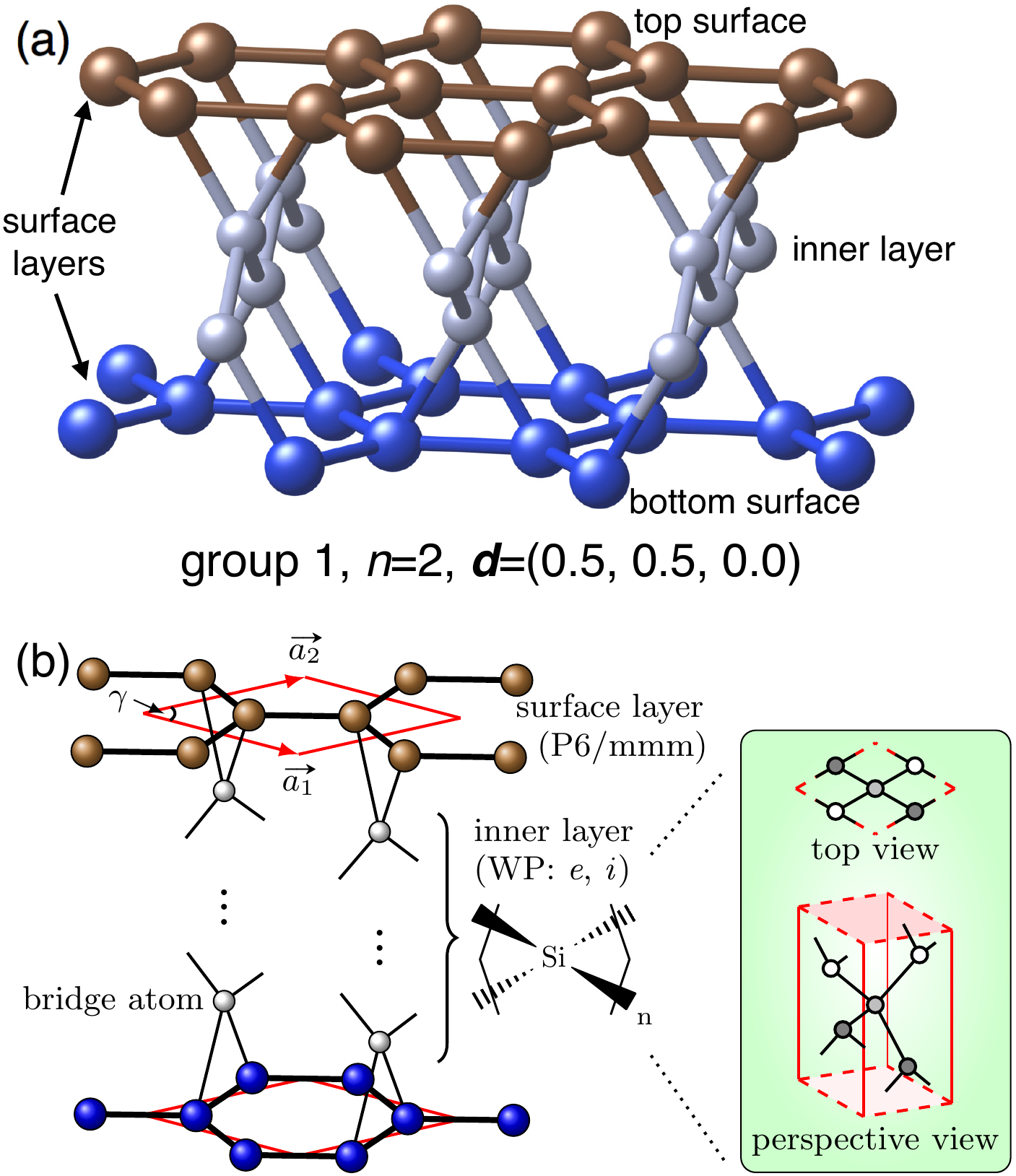}
  \caption{Atomic configuration of the 2D Si crystals. (a) A ball-stick model of the ground state geometry of a 2D Si crystal. Si atoms in top and bottom surface layers are shown in brown and blue balls, respectively, while those in the inner layer, sandwiched by the surface layers, are shown in gray balls. The displacement vector ($\vec{d}$) refers to the relative in-plane displacement of the two surface layers. (b) Schematic diagram of general 2D Si crystals. The unitcell is drawn in red with lattice parameters $\vec{a}_{1}$ and $\vec{a}_{2}$ marked. The inner layer is represented as a distorted tetrahedron, in which filled and dashed wedges by a Cram representation indicate bonds going out of and into the paper, respectively. Top and perspective views of the tetrahedrons in the inner layer are attached on aside, where the atoms in different layers are shaded differently using a gray gradient.}
  \label{fig1}
\end{figure}

The 2D Si crystals constructed by the \texttt{SANDWICH} method demonstrate unique structural features. As we intentionally put together two symmetrically distinct parts (surface and inner layers) so that CN defects on both components are fully compensated by each other, all the atoms in the crystal have a CN of 4 without any dangling bonds as shown in FIG.~\ref{fig1}(a). This condition is particularly favored for Si atoms which show strong preference for sp$^3$ bonding. We find that the four-coordinated networks are maintained in the fully relaxed equilibrium structures. Moreover, the crystals exhibit atomically flat surface structures without buckling or reconstruction, which is uncommon for 2D Si crystals except for silicene bilayers~\cite{bai_graphene-like_2010,aierken_first-principles_2016}. Note that those bilayers with planar surfaces are nothing but the cases in our model with ($\vec{d}$=$\mathbf{0}$, $n$=0)~\cite{bai_graphene-like_2010} and ($\vec{d}$ = 0.5 $\vec{a}_{1}$ + 0.5 $\vec{a}_{2}$, $n$=0)~\cite{aierken_first-principles_2016}.

\begin{figure}[b]
  \centering
  \includegraphics[width=\columnwidth]{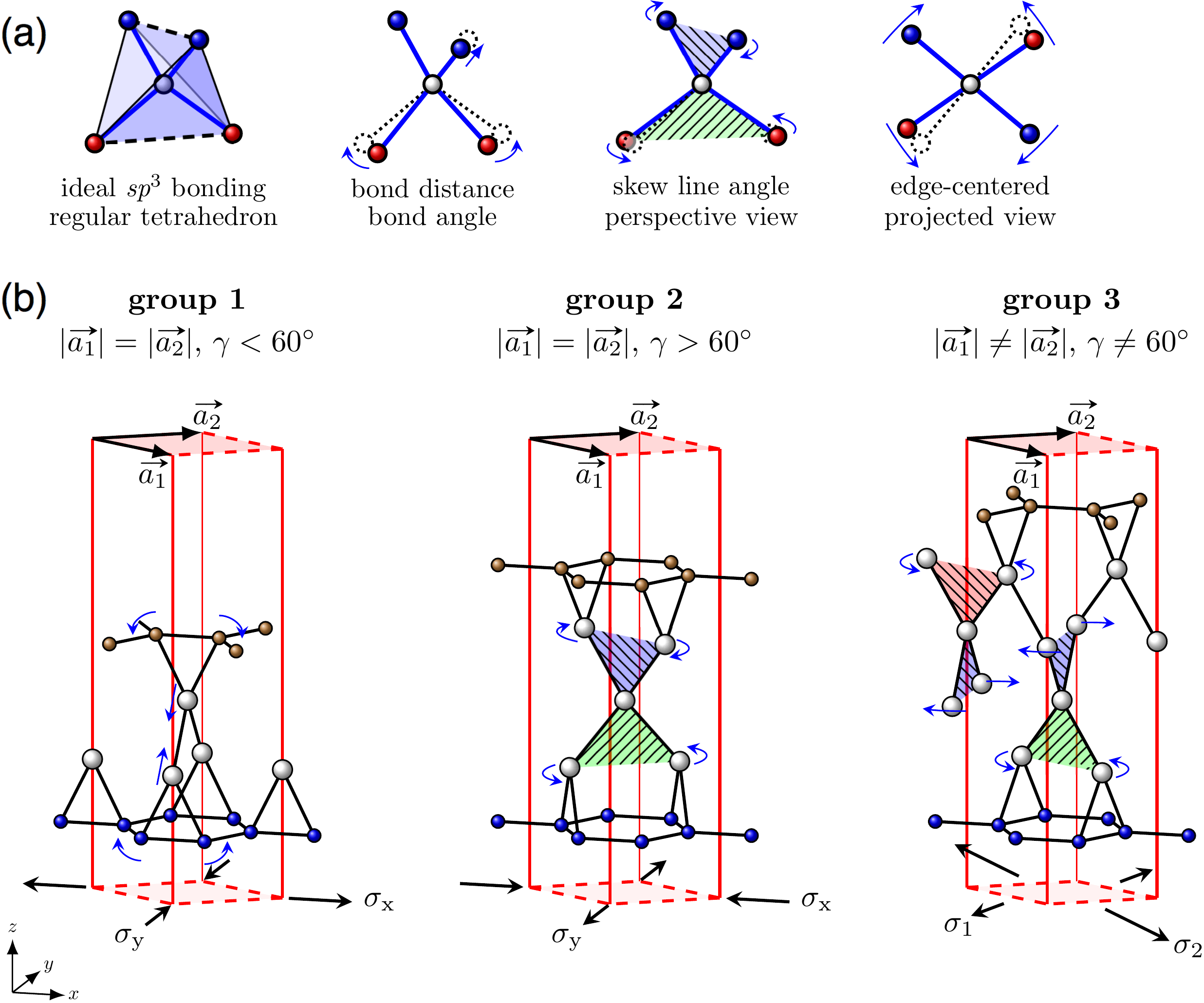}
  \caption{Classification of the 2D Si crystals. (a) Type of distortions in a tetrahedron building block. On the undistorted regular tetrahedron, skew lines along in-plane directions are marked as thick dashed lines. Bond stretch, bending, and twist are shown aside. The undistorted configurations are drawn in dotted lines for comparison, and restoration forces due to the distortion are shown in blue arrows. (b) Classification of the crystals due to various lattice distortions. Local and global stresses are shown in blue and black arrows, respectively. Group 1: local distortions in the surface layers mainly due to bond bending lead to the global stresses of positive (negative) normal stress along x (y) axis. Group 2: the unitcell is deformed by twist-like distortions in the inner layer, resulting in negative (positive) normal stress along x (y) axis. Group 3: dihedral angle distortions from different inner layers yield nonzero shear stress, so that the deformation of the unitcell becomes asymmetric ($|\vec{a}_{1}|\ne|\vec{a}_{2}|$).}
  \label{fig2}
\end{figure}

Looking into the structures in detail, we find that the surface and inner layers have different bonding characteristics as expected from the fact that they initially had different symmetries. The surface layers show distorted hexagonal lattices with additional bonds toward the inner layer, while the atoms in the inner layer (hereafter called as bridge atoms as denoted in FIG.~\ref{fig1}(b)) form distorted tetrahedral bonding with its two opposite edges parallel to the plane of the surface (FIG.~\ref{fig1}). The atomic arrangement of the inner layer is similar to that of the \{100\} surfaces of the cubic diamond phase (dSi) distorted by an in-plane shear strain. We note that the key role of the outermost bridge atom located on the bond center of the surface atoms for stabilizing the characteristic planar surface structure of the crystals, which would not have been realized otherwise. Due to the discrepancy of the preferred local environments for the two parts in the common unitcell, some of the local structures must be distorted for compatibility. For instance, the angle between those bridge atoms and surface atoms is largely deviated from the ideal value of $\sim$109.5 degrees ($\degree$) to $\sim$60$\degree$. Also, the angles between the two skew lines (marked as thick dashed lines in FIG.~\ref{fig2}(a)) are deviated from the right angle (90$\degree$) for some of the tetrahedrons made by bridge atoms, which creates torsional restoration forces as explained in FIG.~\ref{fig2}(a). Among the unique structural features shared by the crystals in this study (namely, flat surfaces without CN defects), we find that a set of 2D crystals constructed with the same $\vec{d}$ and $n$ display a variety of microstructures depending on the atomic arrangement in the inner layer. This comes from the various relative orientations of the distorted sp$^3$ bonds as explained above. 

The global stress of the crystals is also affected critically by the $\vec{d}$, $n$ and the atomic arrangement in the inner layer, making the unitcells distorted in various ways (Table~S1). Thus, for the sake of discussion, we classify the crystals into three distinct groups. (1) The crystals in the first group (group 1) are characterized by the same magnitudes of the lattice vectors ($|\vec{a}_1|$=$|\vec{a}_2|$) and reduced cell angle ($\gamma<60\degree$). In this case, the unitcell is distorted by the normal components ($\sigma_x$ and $\sigma_y$) only, as the shear component ($\tau_{xy}$) is vanished due to symmetry as illustrated in FIG.~\ref{fig2}(b). All the crystals in this group have $\vec{d}$ = 0.5 $\vec{a}_1$ + 0.5 $\vec{a}_2$. Those crystals fall into the C222 space group, except for the case of $n$=2 which is in the Cmme group as in the Table~S1. The magnitude of the global stress seems to be decreasing for thicker crystals (or with the increased $n$), indicated by the $\gamma$ approaching to 60$\degree$. (2) The crystals in the second group (group 2) feature the symmetric lattice vectors ($|\vec{a}_1|=|\vec{a}_2|$) with $\gamma>$60$\degree$. As in the case of group 1, only the $\sigma_x$ and $\sigma_y$ distort the unitcell without shear strain because of the symmetry; the crystals also fall into the space group of C222. Crystals in group 1 and 2 differ by the signs of the normal stresses as indicated in FIG.~\ref{fig2}(b). The crystals in group 2 have two subgroups of $\vec{d}$ = $\mathbf{0}$ and $\vec{d}$ = 0.5 $\vec{a}_1$ + 0.5 $\vec{a}_2$. They behave differently in terms of the $\gamma$ with respect to the $n$ as can be seen in the Table~S1. In the former case, symmetry of the surface layers overtakes that of the inner layer, and the $\gamma$ approaches to 60$\degree$ as the $n$ increases. On the other hand, the crystals with the $\vec{d}$ = 0.5 $\vec{a}_1$ + 0.5 $\vec{a}_2$ subgroup are dominated by the symmetry of the inner layer, making the $\gamma$ approach to 90$\degree$ as the $n$ increases. (3) Lastly, group 3 crystals are constructed with nonzero $\tau_{xy}$, resulting in the asymmetric lattice vectors ($|\vec{a}_1|\ne|\vec{a}_2|$) as well as the space group of P2 with lower symmetry. In this case, the principal stresses ($\sigma_1$ and $\sigma_2$) do not agree with the $\sigma_x$ and $\sigma_y$ due to the $\tau_{xy}$.

\begin{figure}[t]
  \centering
  \includegraphics[width=\columnwidth]{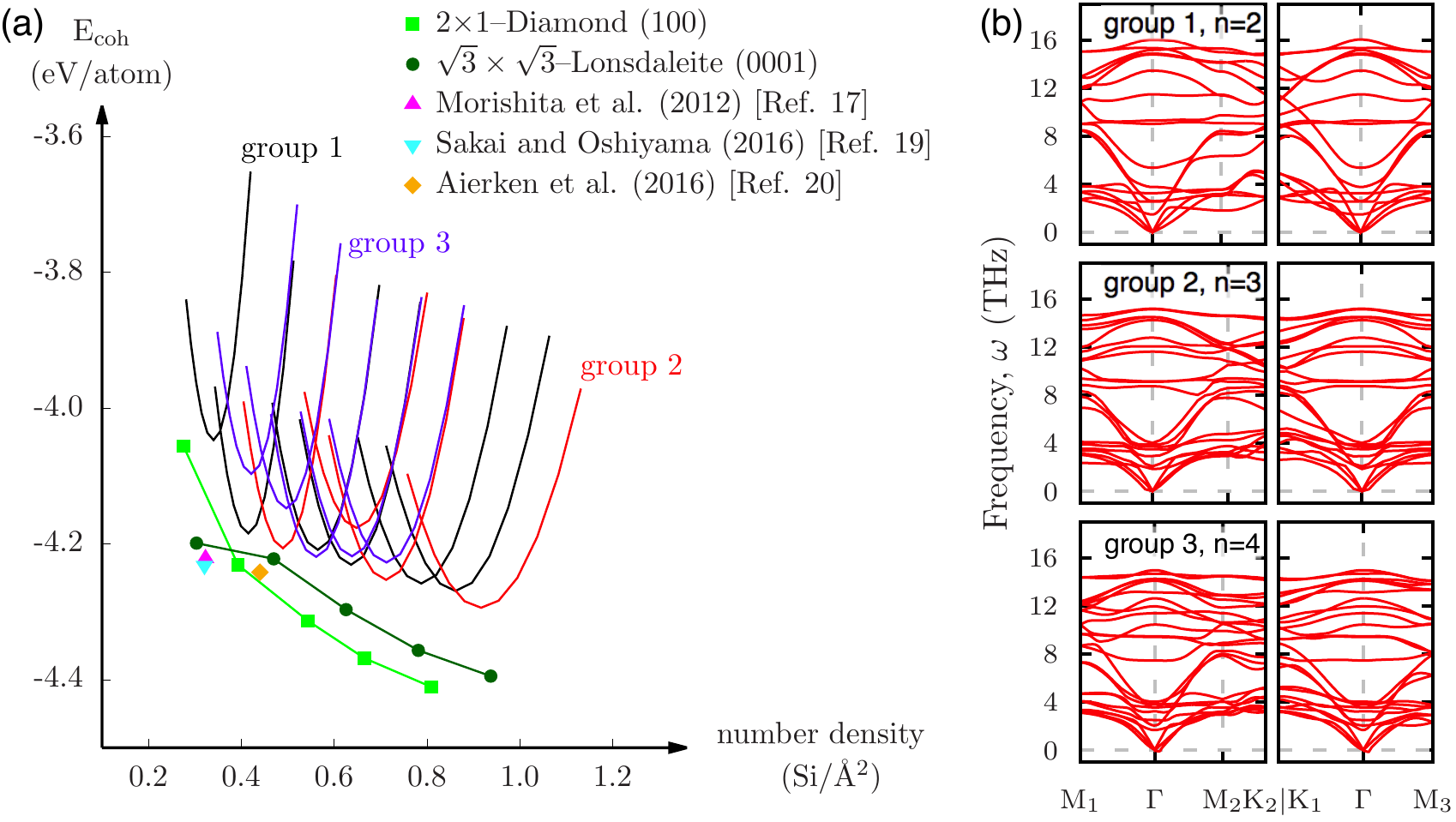}
  \caption{Stability of the crystals. (a) Cohesive energy (E$_{\mathrm{coh}}$) of the 2D crystals with respect to the number density. (b) Harmonic phonon dispersion spectra for a representative structure in each group.}
  \label{fig3}
\end{figure}

Stability of the crystals is confirmed as shown in FIG.~\ref{fig3}. Cohesive energies (E$_{\mathrm{coh}}$) for each crystal are plotted with respect to the number density, defined as the number of Si atoms in the unitcell divided by the unitcell area ($|\vec{a}_1\times\vec{a}_2|$). In all the proposed cases, well-defined energy minima are shown as in FIG.~\ref{fig3}(a). When compared with other Si structures, the crystals in this study show slightly higher E$_{\mathrm{coh}}$. For example, with a number density of $\sim$0.4 Si/\AA$^2$, the crystal with a thickness of 0.5 nm shows a higher E$_{\mathrm{coh}}$ by 0.02 eV/atom than that of 2$\times$1-dSi (100) with a thickness of 0.6 nm. We ascribe this to the significant distortion in bond angle especially near the surface as mentioned above. We note, however, that the 2D crystals in this study may stay more stable than other crystals in a chemical environment because the dangling bonds on the surface were eliminated. This chemical stability undoubtedly becomes a critical factor for device realization~\cite{geim_van_2013,lin_defect_2016}. Moreover, a recently synthesized allotrope of the 3D Si crystals, namely Si$_{24}$, also shows distorted bond angles ranging from 93.73$\degree$ to 123.17$\degree$, and extended bond distances with a higher total energy than that of the ground state dSi by 0.09 eV/Si~\cite{kim_synthesis_2015}. Thus, for realization of stable Si crystals, it could be more critical for the individual Si atoms to satisfy the ground state CN than to maintain the exact bond angles and distances of the ideal sp$^3$ bonding found in dSi. We also provide harmonic phonon dispersion spectra in FIG.~\ref{fig3}(b) for the 2D crystals in the three groups. In all the cases, we confirm that the crystals are dynamically stable.

Furthermore, we find that the new 2D Si crystals are quite stable against strong perturbations beyond a usual harmonic interatomic force regime. To check such a structural stability, we generated computationally ``shaken'' structures that have been proved useful to check the stabilities of other crystal structures~\cite{pickard_ab_2011} (see supplementary material). Starting from the fully relaxed crystal in group 1 with $n$=2, we displaced every Si atom in a random direction with a fixed amount of 0.5~\AA\, in the 2$\times$2 supercell. Note that we used the significantly large displacement when compared with that of 0.01~\AA\, used to obtain harmonic interatomic force constants (FIG.~\ref{fig3}(b)). Then, each of the perturbed structures was relaxed to the corresponding energy minimum configurations by using the conjugate gradient method. By comparing $\sim$5,500 randomly generated configurations, we found that our proposed structure was retained in the $\sim$3,500 samples after relaxation, proving that the structure is robust against severe thermal fluctuations as shown in the FIG.~S2.

\begin{figure}[t]
  \centering
  \includegraphics[width=\columnwidth]{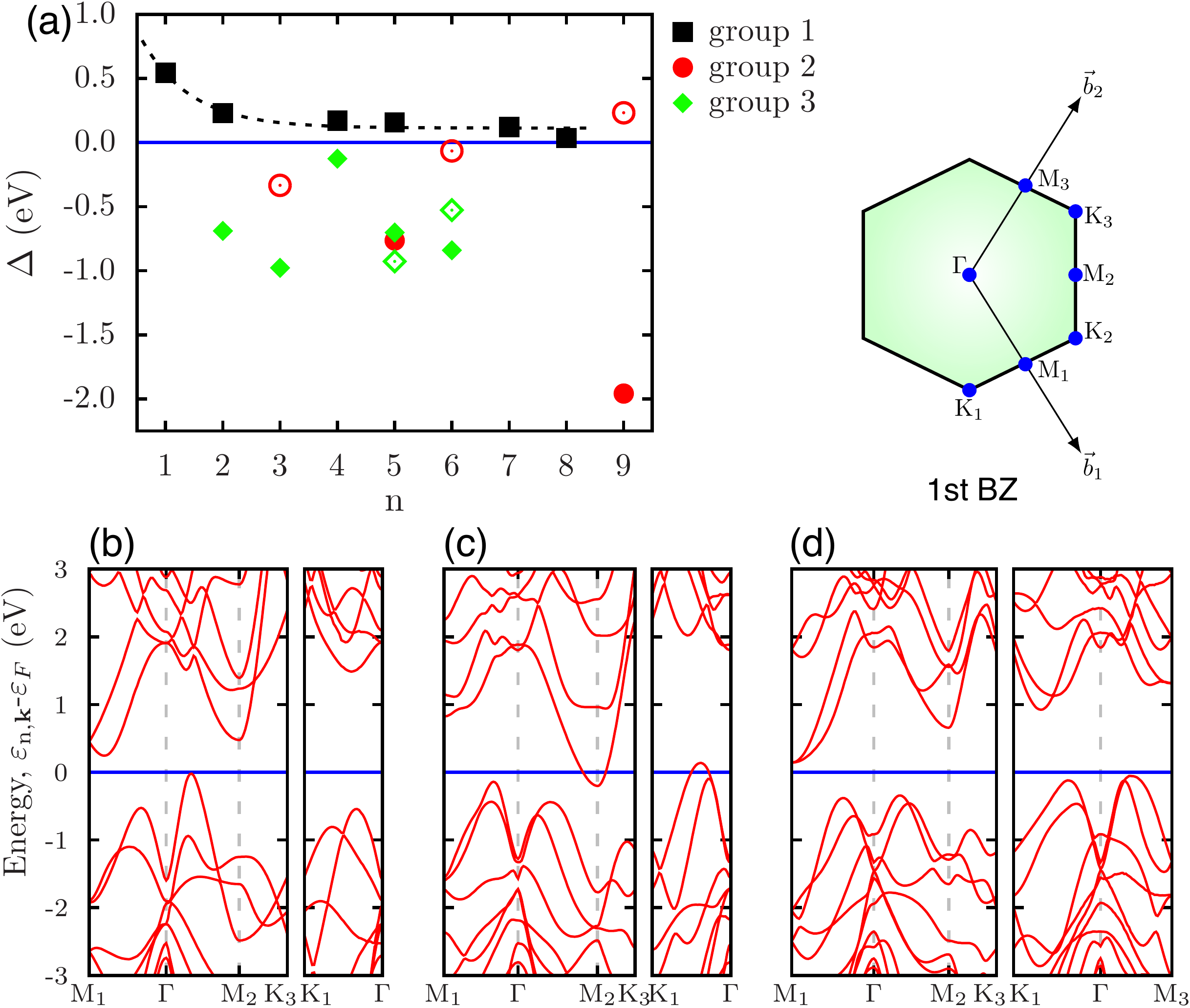}
  \caption{Electronic structures of the 2D crystals. (a) Bandgap ($\Delta$) as a function of thickness ($n$) is shown for the 2D crystals in each group. In group 2 and 3, empty and filled symbols indicate the displacement vectors $\vec{d}$ = $\mathbf{0}$ and $\vec{d}$ = 0.5 $\vec{a}_1$ + 0.5 $\vec{a}_2$. Electronic band dispersion of (b) group 1 with $n$=2, (c) group 2 with $n$=3 and (d) group 3 with $n$=4. A schematic diagram of the 1st BZ with high symmetry points is shown above (d).}
  \label{fig4}
\end{figure}

The fundamental electronic structures of the crystals are closely related to their thickness ($n$) and the crystal classification defined above, displaying a wide variety of electronic properties ranging from metallic to semiconducting. For instance, all the crystals categorized as group 1 are semiconductors, showing a finite bandgap ($\Delta>0$) with their sizes decreasing with an increasing $n$ as shown in FIG.~\ref{fig4}(a). The maximum size of the energy gap is $\sim$0.5 eV for the thinnest crystal ($n$=1). On the other hand, most of the crystals in group 2 and group 3 are metallic. Note that the $\Delta$ in this case is a measure of the overlap between the conduction and valence energy bands. For group 2, the $\Delta$ behaves differently with respect to the $n$ depending on the subgroup ($\vec{d}$) as discussed above.

\begin{figure}[b]
  \centering
  \includegraphics[width=\columnwidth]{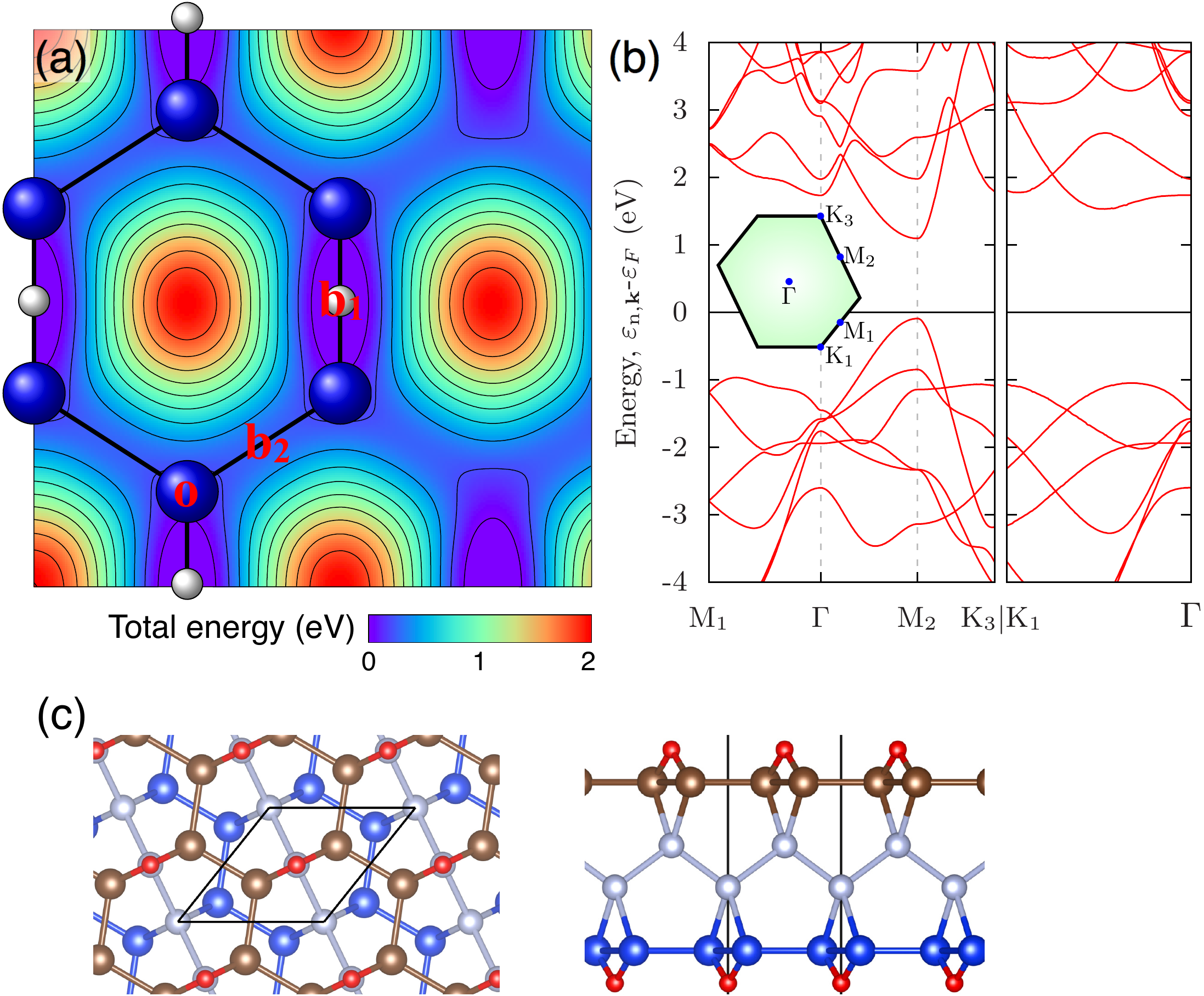}
  \caption{Oxidation of the 2D Si crystal. (a) Potential energy surface of the 2D crystal by an O probe atom with a height of 2 \AA. Atoms are superimposed on the map: large blue and small white balls indicate surface silicon atoms and underneath bridge silicon atoms, respectively. The minimum energy for adsorption site at b$_1$ is set to be zero in the color bar shown below. (b) Electronic band dispersion and (c) atomic structure of the fully oxidized crystal in ground state. Adsorbed oxygen atoms are denoted by small red balls.}
  \label{fig5}
\end{figure}

To reveal the origin of various electronic phases in the crystal families, we provide electronic band dispersions for each group in FIG.~\ref{fig4}(b)-(d). Edges of the valence and the conduction bands are located at different momenta ($\vec{k}$), indicating that the crystals are either indirect semiconductors or semimetals. For the crystals with a space group of C222 (group 1 and 2), we note that the relative energetic positions of the bands at $\vec{k}$ = 0.5 $\vec{b}_1$ ($\epsilon_{M_1}$) and $\vec{k}$ = 0.5 $\vec{b}_1$ + 0.5 $\vec{b}_2$ ($\epsilon_{M_2}$) are directly responsible for the electronic phase transition. Here, $\vec{b}_1$ and $\vec{b}_2$ indicate the reciprocal lattice vectors in the BZ as shown in FIG.~\ref{fig4}. That is to say, when the conduction band minimum (CBM) is located at M$_1$, the crystals represent finite energy gaps ($\Delta>0$), while semimetallic phases with $\Delta<0$ are realized for crystals with the CBM located at M$_2$ as seen in FIG.~\ref{fig4}(b) and (c). Note that the crystals in group 1 and 2 belong to the former and the latter cases, respectively. Because crystals are classified by the lattice parameters of the unitcell (FIG.~\ref{fig2}), we ascribe the electronic phase transition to the changes in the local atomic structures due to the lattice parameters, especially to the $\gamma$. We firstly confirm that the electronic wave functions for the CBM are highly localized near the surface layers (FIG.~S3), indicating that the surface geometry is mainly responsible for the electronic phase transition. We further verify this by pure shear deformation of the unitcell only to change the $\gamma$ without changing $|\vec{a}_1|$ and $|\vec{a}_2|$, and see that the crossover between $\epsilon_{M_1}$ and $\epsilon_{M_2}$ indeed occurs as shown in FIG.~S4. Based on those facts, we can construct a concise but essential model for the whole crystal explaining the characteristic variation of electronic structures from semiconductor to semimetal as functions of thickness and group classification (see FIG.~S5). In addition, these crystals provide good transport properties when compared to the dSi. We note that the transverse effective masses for electrons near the CBM are reduced by half when compared with the dSi, while similar values of other components are shown as in the Table~S2. The structures with lower crystal symmetry (group 3) always show semimetallic electronic structures (FIG.~\ref{fig4}(d)). 

Interestingly, we find that the surface oxidation can extremely widen the applicability of the new crystals. With the oxygen (O) adsorption on the surface, the crystal in group 1 with $n$=2 is significantly stabilized by 1.98 eV/O$_2$, and with the subsequent dissociation of the adsorbed O$_2$ molecule on the surface, the system becomes even more stabilized by 5.89 eV/O$_2$ (See supplementary material for details). We confirm that adsorption and dissociation of O$_2$ molecules do not affect much on the characteristic planar structure of the surface in a wide range of O coverage from an isolated limit (FIG.~S8) to the full coverage (FIG.~\ref{fig5}(c)). We find that the ground state occurs when the adsorbed O atom is located on the middle of the surface Si-Si bond just on top of the underneath bridge Si atom marked as b$_1$ in FIG.~\ref{fig5}(a). Moreover, electronic properties vary notably \emph{via} surface oxidation from the indirect energy gap of $\sim$0.2 eV (FIG. 4(a)) to the direct bandgap of $\sim$1.2 eV (FIG.~\ref{fig5}(b)), which significantly widens the versatility of the 2D Si crystals in this study. Furthermore, we confirm that the oxidized 2D crystals can form stable layered structure by itself, suggesting this Si material as a feasible candidate for a component in vdW heterojunction~\cite{geim_van_2013}.

\section{Conclusions}

In this work, using a newly developed \emph{ab initio} computational method, we propose a series of two-dimensional silicon crystals with versatile electronic properties. The surface layer of the new 2D Si crystals exhibits atomically flat distorted hexagonal structure without buckling, and the inner layer silicon atoms fill up the space between the flat surface layers. We classified 2D Si structures into three groups and each of the groups possesses distinct electronic properties originated from structural variations such as semiconductor as well as semimetals. Moreover, their oxidized forms are shown to be a direct bandgap semiconductor. Therefore, we believe that our new 2D Si crystals satisfy highly desirable characteristics of next generation electronic technology platforms only with a single atomic element and their oxides, very similar with the current 3D Si electronic devices.

\begin{acknowledgements}
The authors thank Dr. In-Ho Lee for discussions. D. Y. K. acknowledges financial support from the NSAF (U1530402). Y.-W. S. was supported by the NRF of Korea funded by the MSIP (QMMRC, No. R11-2008-053-01002-0 and vdWMRC, No. 2017R1A5A1014862). The computing resources were supported by the Center for Advanced Computation of KIAS.
\end{acknowledgements}

%

\widetext
\clearpage

\begin{center}
\textbf{\large Supplementary Material for: \\Atomically flat two-dimensional silicon crystals with versatile electronic properties} \\
\vspace{10pt}
Kisung Chae,$^1$ Duck Young Kim,$^2$ and Young-Woo Son$^1$ \\
\vspace{4pt}
$^1$ \emph{Korea Institute for Advanced Study, Seoul 02455, South Korea} \\
$^2$ \emph{Center for High Pressure Science and Technology Advanced Research, Shanghai 201203, P. R. China} \\
(Dated: \today)
\end{center}

\renewcommand{\thesection}{S\Roman{section}}
\setcounter{section}{0}
\renewcommand{\thefigure}{S\arabic{figure}}
\setcounter{figure}{0}
\renewcommand{\thetable}{S\arabic{table}}
\setcounter{table}{0}
\renewcommand{\bibnumfmt}[1]{[S#1]}
\renewcommand{\citenumfont}[1]{S#1}

\section{A new structure search method for 2D crystal prediction}
\begin{figure}[b]
  \includegraphics[width=6cm]{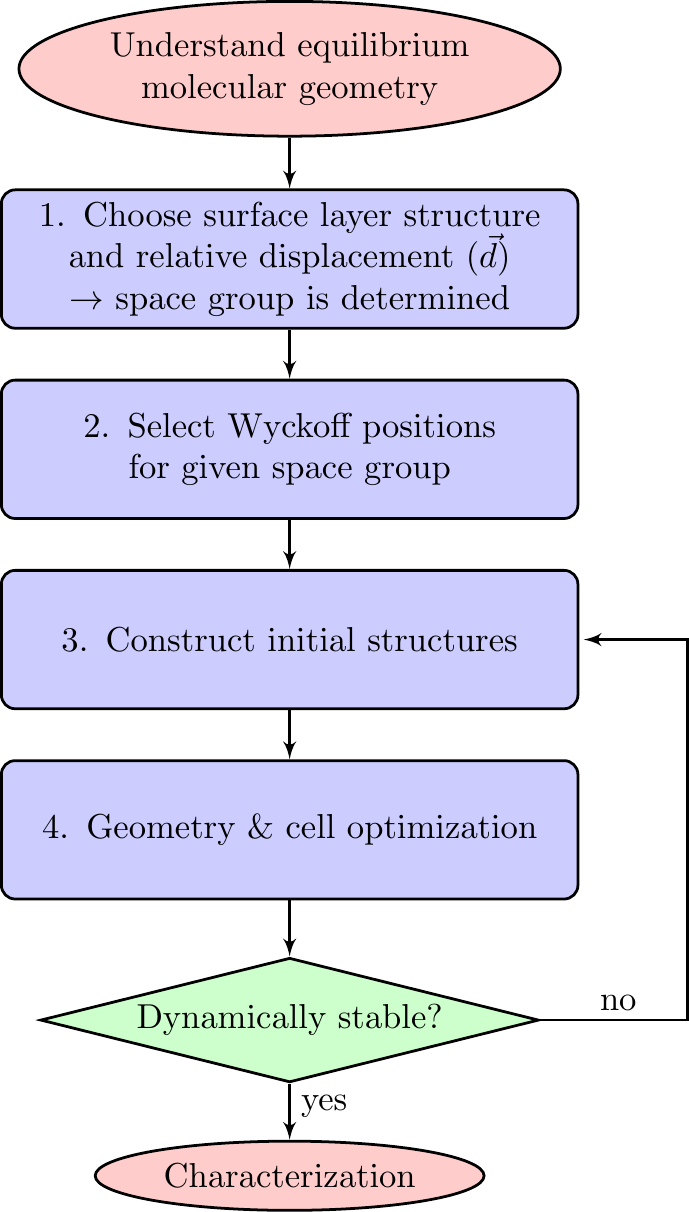}
  \caption{A flowchart of the \texttt{SANDWICH} method.}
  \label{figs1}
\end{figure}

Here, we describe in detail the structure searching method used to predict the 2D Si crystals in this study, named \texttt{SANDWICH} (Search by \emph{Ab initio} Novel Design \emph{via} Wyckoff positions Iteration in Conformational Hypersurface). The main idea of the method is to put together two symmetrically distinctive parts to compensate unpaired electrons at the interface. Thus, the method is particularly suitable for predicting 2D crystals which might favor electronic compensation over local distortion. By doing so, we believe that a new series of 2D crystals can be efficiently searched in a highly confined conformational space near the local minimum structures. The number of distinct sets of crystals that can be constructed by this method can be as many as the possible combinations of surfaces and inner parts, which varies depending on the molecular geometries of the ground state structure (i.e., tetrahedral building block for Si). General procedures of the \texttt{SANDWICH} method are summarized as a flowchart in FIG.~\ref{figs1}, and structural parameters of 2D crystals in this study are listed in the Table~\ref{tab:structure}.

To be more specific, the first step in the \texttt{SANDWICH} method begins with the choice of geometry of surface layers, which determines the space group of the 2D crystals. Note that the two surface layers can be displaced relative to each other by a fractional lattice translation vector $\vec{d}$, and only specific $\vec{d}$ vectors are allowed only when the set of Wyckoff positions in the given space group remain invariant after a translation operation by $\vec{d}$. For instance, the two $\vec{d}$ vectors of $\mathbf{0}$ and 0.5 $\vec{a}_{1}$ + 0.5 $\vec{a}_{2}$ were considered in this study because a set of Wyckoff positions of $e$ (0, 0, $\pm$z) and $i$ [(1/2, 0, $\pm$z), (0, 1/2, $\pm$z) and (1/2, 1/2, $\pm$z), respectively] under the space group of P6/mmm (No. 191) are invariant with respect to both of the $\vec{d}$ vectors. For the same surface layers, if Wyckoff positions of $e$ and $h$ [(1/3, 1/3, $\pm$z) and (2/3, 2/3, $\pm$z)] are chosen to construct the initial structures, there must be three $\vec{d}$ vectors of $\mathbf{0}$ and $\pm$1/3 $\vec{a}_{1}$ + $\pm$1/3 $\vec{a}_{2}$. The next step is to select some of the Wyckoff positions for inner layer construction from the list given for the space group. Here, a rule of thumb is that all the atoms in the constructed crystals have to be free of CN defects. At this point, for a given number of atomic layers in the inner layer n (or thickness), a set of crystals can be constructed. Then, starting with those initial guess structures, the corresponding ground states are sought, followed by dynamic stability tests.

\begin{table}[t]
\caption{Structural parameters of 2D crystals in various groups.}
\begin{tabular}{cccccccc}
\hline
\hline
index & $\vec{d}$ & n & $\vec{a}_1$ & $\vec{a}_2$ & $\gamma$ ($\degree$) & t (\AA) & space group (No.) \\
\hline
\multirow{6}{*}{group 1} & \multirow{6}{*}{(0.5, 0.5)}
   & 1 & 4.165 & 4.165 & 57.94 & 4.93 & C222 (21) \\
 & & 2 & 4.157 & 4.157 & 56.72 & 5.34 & Cmme (67) \\
 & & 4 & 4.086 & 4.086 & 58.04 & 7.89 & C222 (21) \\
 & & 5 & 4.085 & 4.085 & 57.90 & 9.35 & C222 (21) \\
 & & 7 & 4.041 & 4.041 & 58.77 & 12.03 & C222 (21) \\
 & & 8 & 4.035 & 4.035 & 58.79 & 13.42 & C222 (21) \\
\hline
\multirow{5}{*}{group 2} & \multirow{3}{*}{(0.0, 0.0)}
   & 3 & 3.974 & 3.974 & 64.80 & 6.70 & C222 (21) \\
 & & 6 & 3.968 & 3.968 & 63.11 & 10.72 & C222 (21) \\
 & & 9 & 3.967 & 3.967 & 61.95 & 14.78 & C222 (21) \\
\cline{2-8} & \multirow{2}{*}{(0.5, 0.5)}
   & 5 & 3.885 & 3.885 & 69.74 & 9.57 & C222 (21) \\
 & & 9 & 3.793 & 3.793 & 80.28 & 15.24 & C222 (21) \\
\hline
\multirow{7}{*}{group 3} & \multirow{2}{*}{(0.0, 0.0)}
   & 5 & 4.503 & 3.859 & 53.88 & 9.69 & P2 (3) \\
 & & 6 & 3.976 & 4.126 & 58.75 & 10.70 & P2 (3) \\
\cline{2-8} & \multirow{5}{*}{(0.5, 0.5)}
   & 2 & 3.850 & 4.275 & 59.86 & 5.50 & P2 (3) \\
 & & 3 & 3.910 & 3.893 & 67.65 & 6.88 & P2 (3) \\
 & & 4 & 3.880 & 4.360 & 57.41 & 8.13 & P2/c (13) \\
 & & 5 & 3.876 & 4.027 & 64.52 & 9.37 & P2 (3) \\
 & & 6 & 3.879 & 4.561 & 53.18 & 11.00 & P2/c (13) \\
\hline
\hline
\end{tabular}
\label{tab:structure}
\end{table}

We note that some of the 2D Si crystals already reported elsewhere can also be found by the method proposed here. For instance, the crystals, so called a bilayer silicene~\cite{bai_graphene-like_2010s,aierken_first-principles_2016s}, in AA (ontop) and orthorhombic (displaced along the zigzag chain direction) stacking of the two silicene monolayers are nothing but the ones with ($\vec{d}$=0, n=0) and ($\vec{d}$=0.5 $\vec{a}_{1}$ + 0.5 $\vec{a}_{2}$, n=0), respectively. Similarly, it is theoretically shown that other group IV elements such as germanium (Ge) and tin (Sn) with the same valence electron configuration as Si can form similar bilayer structures as well~\cite{acun_germanene:_2015s,huang_quantum_2016s}.

\section{Structural stability test}

\begin{figure}[t]
  \includegraphics[width=8cm]{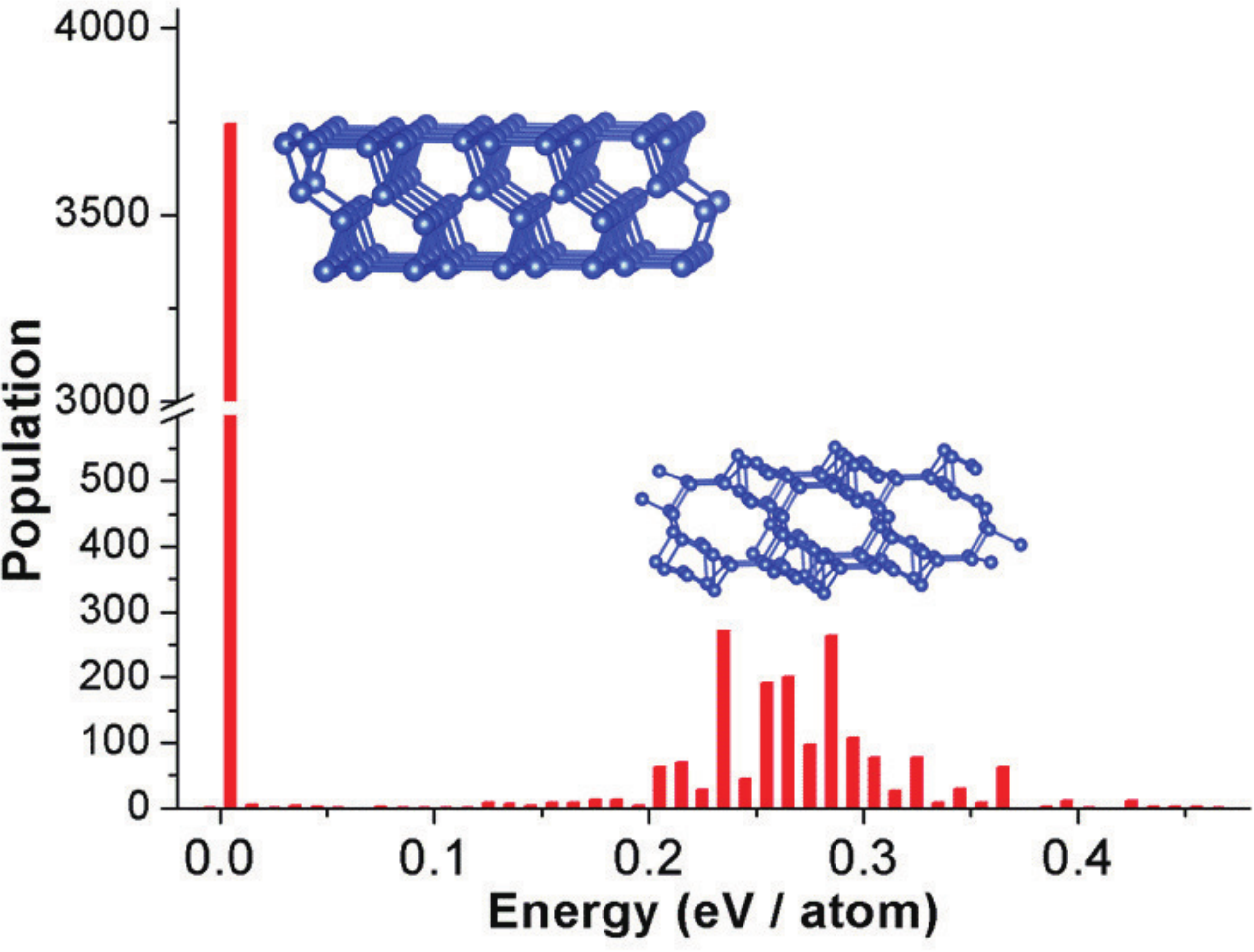}
  \caption{Histogram of total energies of the relaxed structures from stochastic displacements referenced to the lowest energy.}
  \label{figs2}
\end{figure}

In addition to calculating harmonic phonon dispersion to check the dynamic stability, we also tested the stability beyond a harmonic regime. We started from one of our proposed 2D silicon crystals (group 1, n=2), and then randomly moved each atomic position from the equilibrium within a pre-determined spherical space with a fixed radius. This method was used to predict a low energy models of SiNF~\cite{le_page_low_2006s}, and it is one important operation in \emph{ab initio} random structure searching~\cite{pickard_ab_2011s} strategy, which have been successfully applied to many compounds under pressure. We moved every Si atom in a (2$\times$2) supercell by 0.5 \AA, and then allowed a full relaxation. Note that the magnitude of the displacement is significantly larger than that used in harmonic limit of 0.01 \AA. This indicates that the crystals were tested on a very harsh condition or very high temperature. We examined $\sim$5,500 structures and the results are shown in FIG.~\ref{figs2}. We find that majority of relaxed structures after random distortion are recovered back to the original structure, which reaches $\sim$68 \% out of the total samples, and possess the lowest total energy. Rest of the distorted structures were relaxed to local minimum structures with buckled surfaces with higher total energy as shown in FIG.~\ref{figs2}.

\section{Origin of a variety of electronic properties}
\begin{figure}[b]
  \includegraphics[width=4cm]{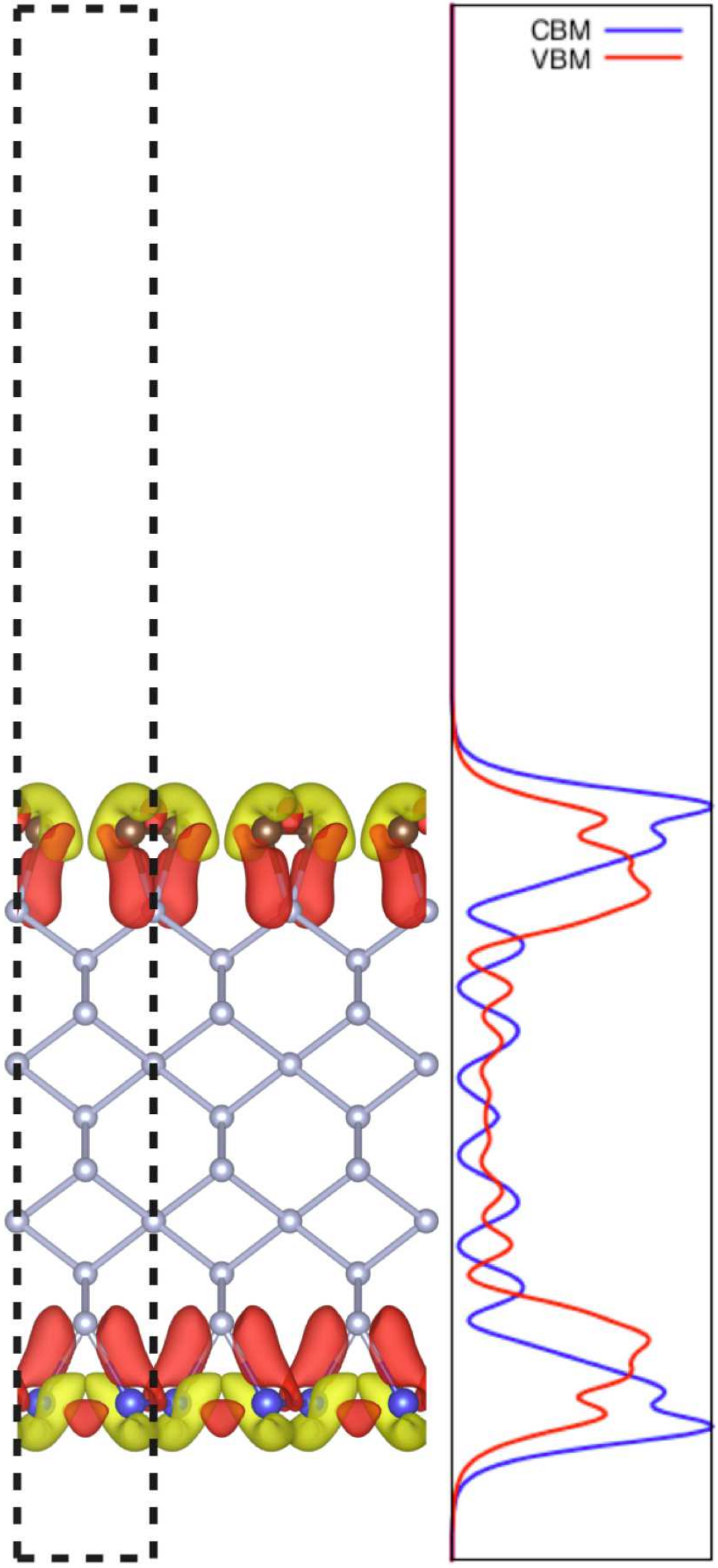}
  \caption{Charge densities for a 2D crystal in group 2 with n=9 at valence band maximum (VBM) and conduction band minimum (CBM) rendered in red and yellow, respectively. An isovalue of 0.005 electron per \AA$^3$ was used. Unitcell is drawn as black dashed line. Charge density projected along the normal direction to the crystal is plotted for CBM and VBM.}
  \label{figs3}
\end{figure}

\begin{figure}[t]
  \includegraphics[width=12cm]{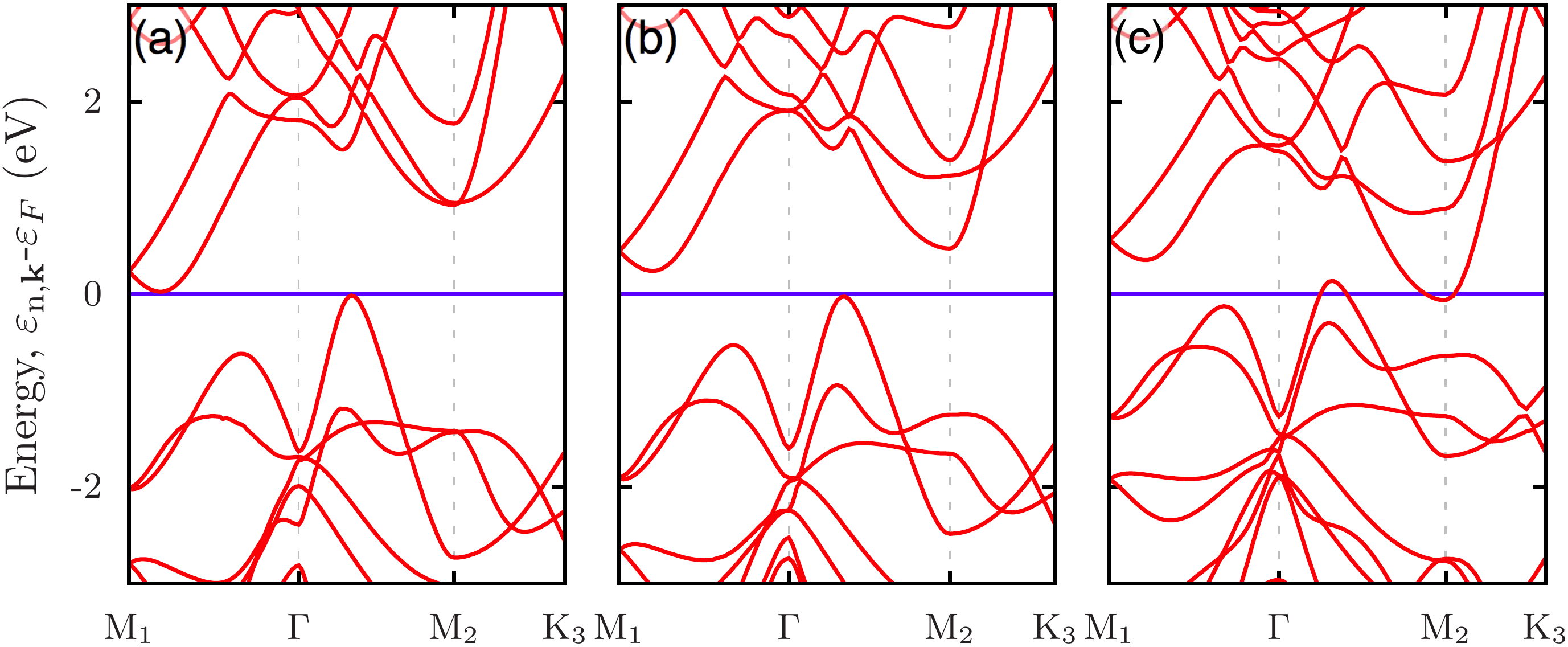}
  \caption{Electronic band dispersion of a 2D Si crystal in group 1 with n=2 as a function of $\gamma$: (a) 54.45$\degree$, (b) 56.72$\degree$ and (c) 58.99$\degree$.}
  \label{figs4}
\end{figure}

We demonstrate that the versatile electronic properties (FIG.~4) are primarily attributed to the surface layer structure. Simple shear deformation, together with the fact that electronic wave functions near the Fermi energy are localized on the surface layers (FIG.~\ref{figs3}), shows that the variation of electronic properties is solely due to the surface geometry, and rules out the role of inner layer (FIG.~\ref{figs4}). Specifically, shift of the band at M$_2$ ($\vec{k}_{M_2}$ = 0.5 $\vec{b}_{1}$ + 0.5 $\vec{b}_{2}$) with respect to the shear strain (or the cell angle $\gamma$) is clearly shown, altering the electronic structures ranging from semimetallic to indirect semiconducting.

\begin{figure}[b]
  \includegraphics[width=\textwidth]{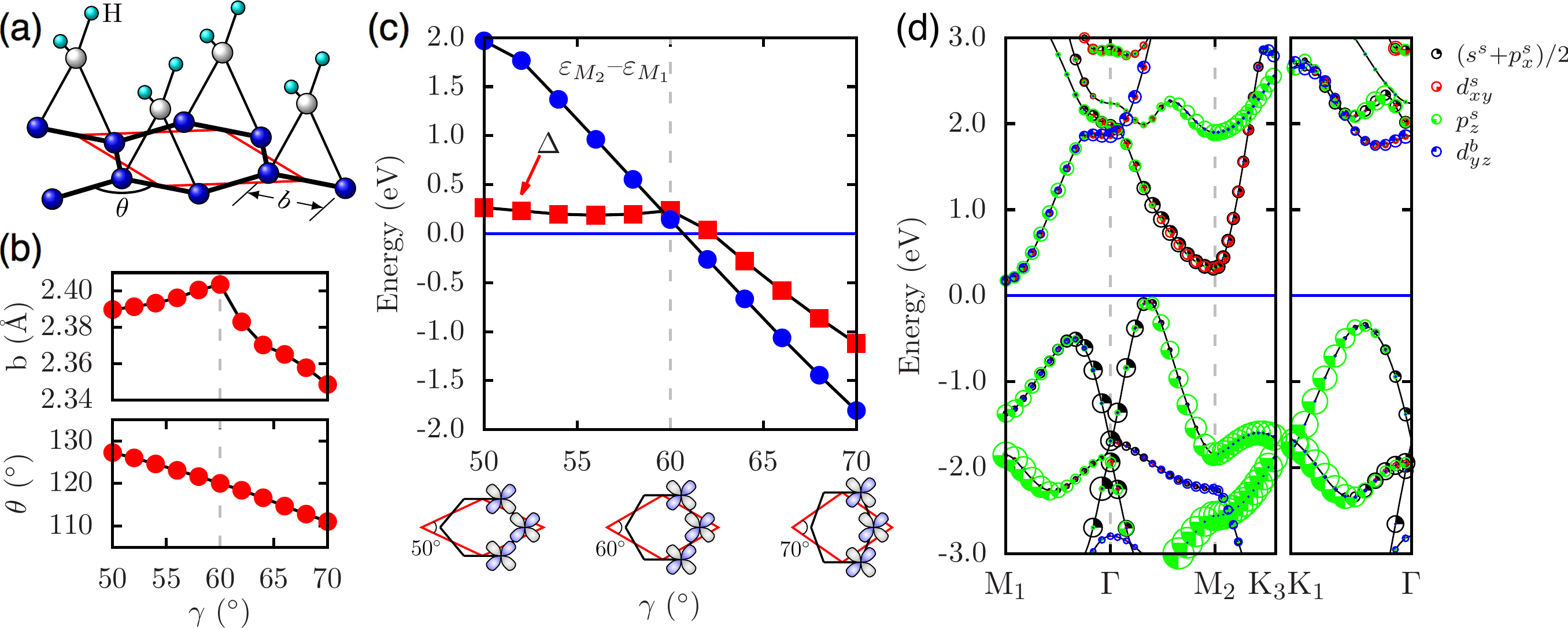}
  \caption{(a) Atomic configuration of the model. (b) Bond length and bond angle marked in (a) with respect to $\gamma$. (c) $\Delta$ and energy level difference at M$_1$ and M$_2$ due to $\gamma$ are shown. The d$_{xy}$ orbital overlaid on the surface layer with a corresponding angle is shown below. (d) Dispersion of the electronic band structure with atomic orbital projection. Orbitals for surface and bridge atoms are denoted as s and b, respectively, on the superscript. Size of the symbols represents the amount of contribution.}
  \label{figs5}
\end{figure}

To reveal the mechanism of the band shift at M$_2$, we consider a minimal surface model. The model is composed of essential parts: a hexagonal framework of Si with a bridge Si atom with hydrogen passivation as shown in FIG.~\ref{figs5}(a). Based on the fact that the band shift is primarily due to the surface layer structure, we varied the $\gamma$ from 50$\degree$ to 70$\degree$. The length of the surface bond with the bridge atom (denoted as b in FIG.~\ref{figs5}(a)) does not vary monotonously with respect to the $\gamma$, but becomes diminished as the unitcell is distorted as seen in FIG.~\ref{figs5}(b). Instead, the bond angle at the surface ($\theta$) decreases monotonously with an increasing $\gamma$, so does the relative band shift between M$_1$ and M$_2$ ($\epsilon_{M_2}-\epsilon_{M_1}$) as seen in FIG.~\ref{figs5}(c). So, the $\theta$ appears mainly responsible for the band shift.

For more rigorous discussion, we show the electronic band structures projected on atomic orbitals of the minimal model in FIG.~\ref{figs5}(d). The band near the valence band maximum seems to be composed mainly of the s, p$_x$ and p$_z$ states of the surface atoms. On the other hand, orbital characters of the band at M$_1$ and M$_2$ vary significantly. The band at M$_1$ contains a significant portion of atomic orbital perpendicular to the surface layer such as p$_z$, d$_{xz}$ and d$_{yz}$. On the other hand, the band at M$_2$ is mainly composed of orbital lying on the surface layer such as s, p$_x$ and d$_{xy}$. We note that d$_{xy}$ orbital contributes ~22 \% to the band at M$_2$. Thus, we find out that the sensitive band shift at M$_2$ is partly due to the enhanced (diminished) contribution of d$_{xy}$ orbital with the decreased (increased) $\gamma$ as shown in FIG.~\ref{figs5}(c). The bands at M$_1$ and M$_3$ do not vary as much as M$_2$ because the bridge bond varies less sensitively to the $\gamma$, compared to the surface bonds. Therefore, the $\Delta$ remains almost constant even with the $\gamma$ smaller than 60$\degree$.

\section{Effective masses}

\begin{figure}[h]
  \includegraphics[width=8cm]{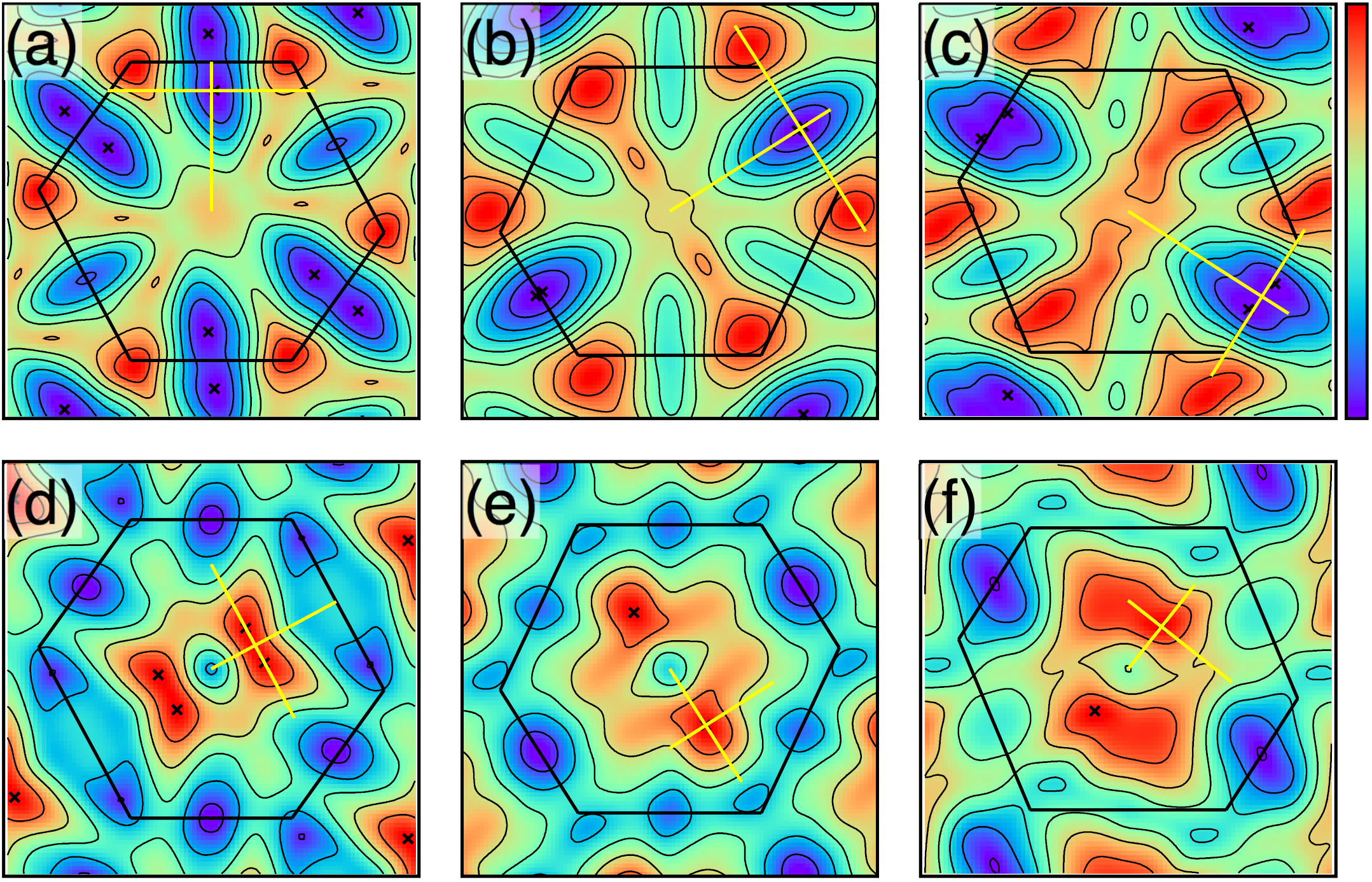}
  \caption{Eigenvalue maps in the reciprocal space. Top and bottom panes indicate conduction and valence bands, respectively, for (a, d) group 1, n=2, (b, e) group 2, n=3, and (c, f) group 3, n=4. Band extrema are marked as x, and the first BZ is drawn as black lines. Effective masses around the band extrema were calculated along longitudinal and transverse direction as marked by dissecting yellow lines.}
  \label{figs6}
\end{figure}

Effective masses for electron and hole are calculated by interpolating energy band dispersion to a parabolic band around the band extrema for selected 2D Si crystals. In general, the 2D crystals in this study have elliptical Fermi surfaces, indicating anisotropic effective masses as seen in FIG.~\ref{figs6}. The calculated effective masses are summarized in the Table~\ref{tab:effmass}. Compared to cubic diamond phase (dSi) with longitudinal ($m^{*}_{e,L}$) and transverse ($m^{*}_{e,T}$) electronic effective masses of 0.92 $m_0$ and 0.19 $m_0$ ($m_0$: mass of a free electron), respectively~\cite{kittel_introduction_2005s}, our predicted crystals in group 1 and 2 show 50 \% lighter $m^{*}_{e,L}$. For hole effective masses, our results are comparable to the dSi: 0.49 $m_0$ and 0.16 $m_0$ for heavy and light holes, respectively. We note that the crystal in group 1 with n=2 can be a good n-type semiconductor with high mobility due to reduced effective mass.

\begin{table}[h]
\caption{Electron (e) and hole (h) effective masses (m$^*$) of the 2D Si crystals with varying group and n. Longitudinal (L) and transverse (T) directions for each case are shown in FIG.~\ref{figs6}.}
\begin{tabular}{ccccc}
\hline
\hline
group, n & m$^*_{e, L}$ & m$^*_{e, T}$ & m$^*_{h, L}$ & m$^*_{h, T}$ \\
\hline
1, 2 & 0.45 m$_0$ & 0.16 m$_0$ & 0.19 m$_0$ & 0.51 m$_0$ \\
2, 3 & 0.52 m$_0$ & 0.18 m$_0$ & 0.43 m$_0$ & 0.22 m$_0$ \\
3, 4 & 1.13 m$_0$ & 0.16 m$_0$ & 0.23 m$_0$ & 0.20 m$_0$ \\
\hline
\hline
\end{tabular}
\label{tab:effmass}
\end{table}

\section{Oxygen adsorption}

\begin{figure}[h]
  \includegraphics[width=3.5cm]{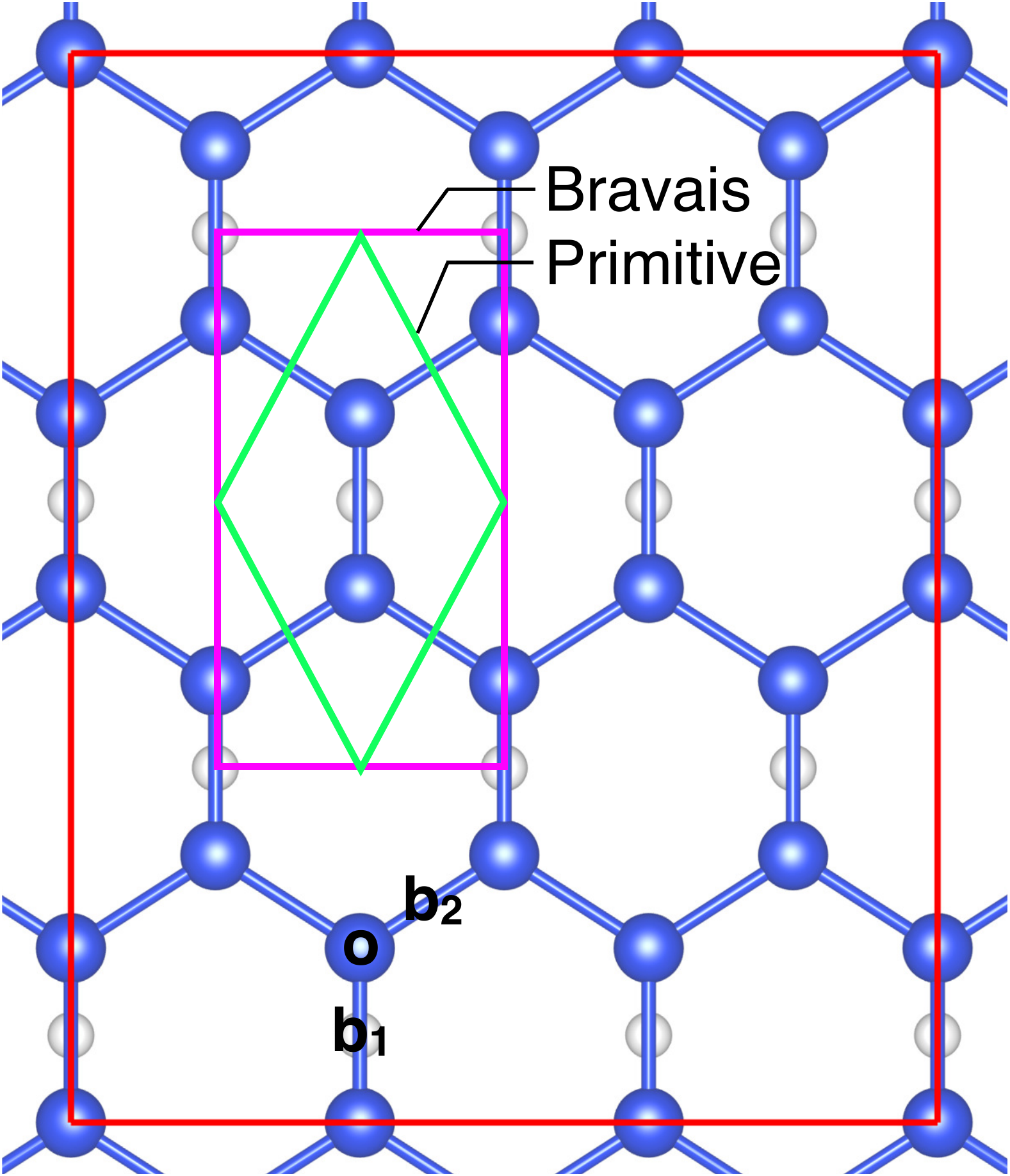}
  \caption{A xy-projection of (3$\times$2) supercell (red) of the conventional Bravais unit cell shown in magenta lines. Only top surface layer (blue balls) and the subsurface bridge atoms (white) are represented. Three possible adsorption sites of ontop (o), bridge 1 (b$_1$) and bridge 2 (b$_2$) are marked on the corresponding sites.}
  \label{figs7}
\end{figure}

In general, Si surface is vulnerable to an oxygen (O) ambient environment due to strong interaction between Si and O, forming a stable oxide film. In addition to their mechanical and dynamical stability as confirmed in FIG.~3, our crystals are expected to show a better chemical stability when compared with other Si crystals, because all the surface Si atoms are fully compensated. To validate this, we attached an isolated O$_2$ molecule on a (3$\times$2) supercell of an orthogonal unitcell as seen in FIG.~\ref{figs7}. As a diatomic molecule, three orientations for O$_2$ alignment along x, y and z were considered for adsorption on the following sites: ontop (o), bridge 1 (b$_1$) and bridge 2 (b$_2$) (FIG.~\ref{figs7}). Note that b$_1$ and b$_2$ are distinguished whether the bridge bond possesses a bridge Si atom on the opposite side or not.

\begin{figure}[h]
  \includegraphics[width=8cm]{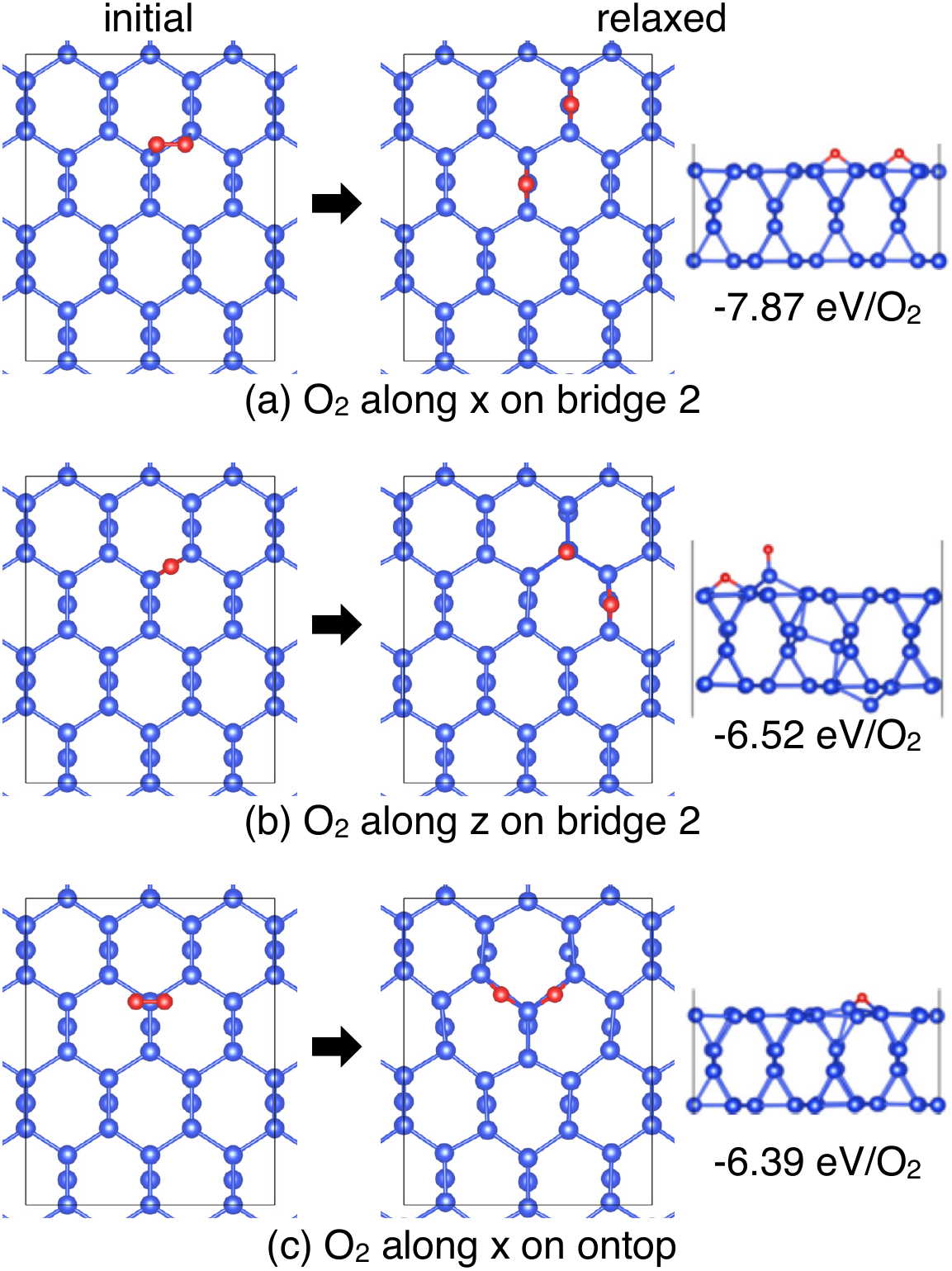}
  \caption{Initial and relaxed structures of the three low-energy configurations of O$_2$ (red balls) adsorption.}
  \label{figs8}
\end{figure}

We observe negative adsorption energies ($E_{\mathrm{ads}}$) as
\begin{equation}
	E_{\mathrm{ads}}=E(\mathrm{Si}+\mathrm{O_2})-E(\mathrm{Si})-E(\mathrm{O_2}),
\end{equation}
indicating that the O$_2$ adsorption is a thermodynamically spontaneous process. Moreover, we observe that internal energy of the system is further reduced in a great deal by dissociation of the adsorbed O$_2$ molecule. Interestingly, the characteristic flat surface of the crystals is retained in most cases throughout adsorption and subsequent dissociation processes (FIG.~\ref{figs8}). This indicates that an oxidation process would make the crystals even more stable without disturbing the original framework, and prevent the structure from being degraded by further oxidation.

In some cases, we observe a significant deformation of the crystal when an O$_2$ molecule is adsorbed on the o site as shown in FIG.~\ref{figs8}(b). We found that the one of the dissociated O atoms is bound to the Si atom, of which the bond orientation is perpendicular to the surface. This changes the local bonding character of the Si to sp$^3$ similar to the dSi (111) surface, so that the flat surface is not preserved anymore. Also, the bridge Si atom, which was originally bound to that Si atom is significantly relaxed, making a new bond with another bridge Si atom. Consequently, the surface Si atom on the other side becomes protruded as seen in FIG.~\ref{figs8}(b). This process seems irreversible because of the strong Si-O bonding, and the protruded Si atom on the other side is likely to be chemically reactive since one of the valence electrons is not fully compensated anymore. However, this process is energetically less favorable by 1.35 eV per a O$_2$ molecule (Table~\ref{tab:Oads}), and O atoms will sit on the most stable b$_1$ site without disturbing the framework in a quasi-static oxidation condition.

\begin{table}[h]
\caption{Adsorption energy ($E_{\mathrm{ads}}$) of an isolated O$_2$ molecule on various adsorption sites.}
\begin{tabular}{cccc}
\hline
\hline
adsorption & initial O$_2$ & E$_{\mathrm{ads}}$ & O$_2$ dissociation \\
site & orientation & (eV/O$_2$) & \\
\hline
\multirow{3}{*}{ontop}
 & x & -6.39 & yes \\
 & y & -4.92 & yes \\
 & z & -2.67 & no \\
\hline
\multirow{3}{*}{bridge 1}
 & x & -4.62 & yes \\
 & y & -2.67 & no \\
 & z & -1.37 & no \\
\hline
\multirow{3}{*}{bridge 2}
 & x & -7.87 & yes \\
 & y & -7.87 & yes \\
 & z & -6.52 & yes \\
\hline
\hline
\end{tabular}
\label{tab:Oads}
\end{table}

\end{document}